\documentclass[journal]{IEEEtran}
\usepackage{graphicx}

\begin{document}
\title{Balloon-borne hard X-ray polarimetry with PoGOLite}
%
%
\author{Mark Pearce, Hans-Gustav Flor\'{e}n, Miranda Jackson, Tune Kamae, M\'{o}zsi Kiss, Merlin Kole, Elena Moretti, G\"{o}ran Olofsson, Stefan~Rydstr\"{o}m, Jan-Erik Str\"{o}mberg, Hiromitsu Takahashi (on behalf of The PoGOLite Collaboration)
\thanks{Manuscript received November 15, 2012. This work was supported in part by The Knut and Alice Wallenberg Foundation, The Swedish National Space Board, The Swedish Research Council and the G\"{o}ran Gustafsson Foundation.}
\thanks{M.~Pearce is with KTH Royal Institute of Technology, Department of Physics, and the Oskar Klein Centre for Cosmoparticle Physics, AlbaNova University Centre, 10691 Stockholm, Sweden (e-mail: pearce@kth.se).}
\thanks{M.~Jackson, M.~Kiss, M.~Kole, E.~Moretti and S.~Rydstr\"{o}m are with KTH Royal Institute of Technology, Department of Physics, and the Oskar Klein Centre for Cosmoparticle Physics, AlbaNova University Centre, 10691 Stockholm, Sweden.}
\thanks{H.-G.~Flor\'{e}n and G.~Olofsson are with Department of Astronomy, Stockholm University AlbaNova University Centre, 10691 Stockholm, Sweden.}
\thanks{J.-E.~Str\"{o}mberg is with DST Control, Link\"{o}ping, Sweden.}
\thanks{H.~Takahashi is with Department of Physical Sciences, Hiroshima University, Hiroshima, Japan.}
\thanks{T.~Kamae is with KiPAC, Stanford University, Menlo Park, USA.}}

\maketitle
\pagestyle{empty}
\thispagestyle{empty}

\begin{abstract}
PoGOLite is a hard X-ray polarimeter operating in the 25-100~keV energy band. The instrument design is optimised for the observation of compact astrophysical sources. Observations are conducted from a stabilised stratospheric balloon platform at an altitude of approximately 40~km. The primary targets for first balloon flights of a reduced effective area instrument are the Crab and Cygnus-X1. The polarisation of incoming photons is determined using coincident Compton scattering and photo-absorption events reconstructed in an array of plastic scintillator detector cells surrounded by a bismuth germanate oxide (BGO) side anticoincidence shield and a polyethylene neutron shield. A custom attitude control system keeps the polarimeter field-of-view aligned to targets of interest, compensating for sidereal motion and perturbations such as torsional forces in the balloon rigging. An overview of the PoGOLite project is presented and the outcome of the ill-fated maiden balloon flight is discussed.   
\end{abstract}

\begin{IEEEkeywords}
polarisation, astrophysics, Compton, Crab, Cygnus X-1, scientific ballooning 
\end{IEEEkeywords}

\section{Introduction}

\IEEEPARstart{I}{t} is 50~years since Giaconni's discovery of extrasolar X-rays. Geiger counters carried outside of the Earth's atmosphere by an Aerobee sounding rocket registered X-rays from Scorpius-X1, now known to be an accreting neutron star. A significant and unexpected diffuse X-ray background was also observed. Since then, X-ray emissions from a plethora of objects have been recorded and our understanding of high-energy processes in the Universe has improved significantly. The resulting field of X-ray astrophysics has been driven by the development of new experimental techniques and, in particular, new methods for detecting X-rays. Giaconni's Geiger counters were subsequently replaced by energy sensitive proportional counters which in turn have been replaced by semiconductor detectors such as charge coupled devices. Instruments are now-a-days flown on satellite missions giving longer observing times for larger collecting areas. A common feature of essentially all missions to date is that they are designed to measure three aspects of the incident X-ray flux, namely the location of origin on the sky, the energy and the arrival time. The majority of instruments are not optimised for measurements of the polarisation of the incident flux. 

Polarised X-rays are expected from the high-energy processes at work within compact astrophysical objects such as pulsars, accreting black holes and jet-dominated active galaxies~\cite{overview}. Polarisation is also expected to provide valuable information regarding the transient processes underlying solar flares~\cite{solar} and gamma-ray bursts (GRB)~\cite{grb}. Polarisation arises naturally for synchrotron radiation in large-scale ordered magnetic fields and for photons propagating through a strong magnetic field. Polarisation can also result from anisotropic Compton scattering. In all cases, the orientation ('angle') of the polarisation plane and degree of polarisation are a powerful probe of the physical environment around compact astrophysical sources. 

\begin{figure}[!t]
\centering
\includegraphics[width=9cm]{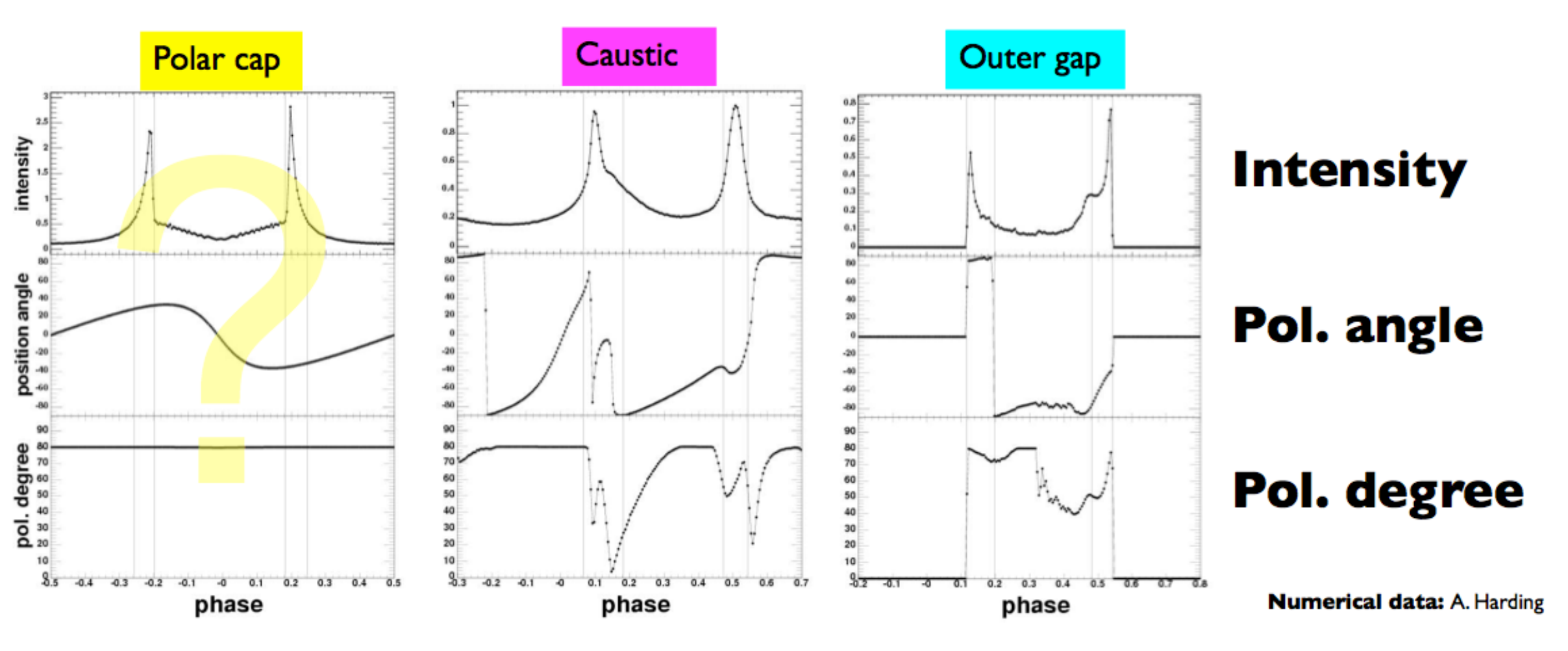}
\caption{Three models proposed to explain high energy emission from the Crab pulsar. The light curve predictions are very similar whereas the polarisation signatures (angle and degree) vary significantly between models.}
\label{fig:Crab}
\end{figure}

The diagnostic worth of polarisation is exemplified for the case of pulsars. Although discovered in 1967, there is still no generally accepted model to explain the high energy emission. In figure~\ref{fig:Crab}, three competing models for high energy emission from the Crab pulsar are shown. In the polar cap model\footnote{It is noted that this model is disfavoured by precision measurements of the Crab emission spectrum by the Fermi Gamma-ray Space Telescope~\cite{Fermi}.}, electrons (positrons) are accelerated in open field line regions at the polar cap regions of the neutron star. The accelerated particles and associated electromagnetic cascades emit synchrotron and curvature radiation while in the vicinity of the non-uniform magnetic field in the polar region. In the caustic model, particle acceleration and emission occur along the edge of the open field region - extending from the surface of the neutron star to the light cylinder. In the outer gap model, electrons are accelerated in the vacuum gaps between the open and closed magnetic field lines in the outer magnetosphere. As shown in the upper panels of the figure, the predicted light-curves for the three emission models are similar - especially when instrument response is included. The polarisation degree (lower panels) and angle (middle panels) show very different signatures for the three models.    

Despite the wealth of sources accessible to polarisation measurements and the importance of these measurements, there has been only one successful mission with dedicated instrumentation optimised for compact sources. The Crab nebula was studied using a Bragg reflectometer flown on the OSO-8 satellite in 1976~\cite{weisskopf}. The polarisation degree at the allowed measurement energies of 2.6~keV and~5.2 keV were found to be (19.2$\pm$1.0)\% and (19.5$\pm$2.8)\%, respectively. The corresponding polarisation angles were (156.4$\pm$1.4)$^\circ$ and (152.4$\pm$4.0)$^\circ$, respectively. The results are consistent with synchrotron emission processes. The polarimeter did not have sufficient effective area to allow significant measurements of the Crab pulsar.

Measurements making inventive use of the IBIS and SPI instruments on-board the INTEGRAL satellite have reinvigorated the field of late, with polarisation measurements and limits reported for both the Crab and Cygnus X-1~\cite{crabIBIS,crabSPI,CygIBIS}. The IBIS instrument was used to observe the Crab in the 200-800~keV range (2003-2007) and a high degree of polarisation, $>$ 72\%, was observed in the region between the emission peaks. As for OSO-8, the high polarisation degree indicates synchrotron emission from a well-ordered magnetic field region. No significant polarisation was found in the peaks. This may be due to a rapid change in the polarisation angle combined with poor timing resolution. The polarisation angle in the off-peak region was found to be (122.0$\pm$7.7)$^\circ$, i.e. consistent with the pulsar rotation axis. The SPI instrument was used to measure polarisation in the 100 keV - 1 MeV energy band (2003-2006). The time averaged polarisation degree was found to be (46$\pm$10)\% for an angle of (123.0$\pm$11)$^\circ$, i.e. consistent with IBIS. The IBIS instrument was also used to study emission from Cygnus X-1. An upper limit of 20\% was placed on the energy band 250-400~keV while a polarisation degree of (67$\pm$30)\% was determined for higher energies up to 2 MeV for an angle of (140$\pm$15)$^\circ$. This is at least 100$^\circ$ from the compact radio jet. Spectral modelling of the data suggests that the low energy measurements are consistent with emission dominated by Compton scattering on thermal electrons. The higher energy data which is, in contrast, strongly polarised may indicate that emission in the MeV region is related to the radio jet. 
It is important to note that the IBIS and SPI instruments were not designed for polarimetric measurements and that their response to polarised radiation was not studied prior to launch. 

The GAP wide field-of-view hard X-ray polarimeter (50-300~keV) was launched on-board the IKAROS solar sail demonstrator in May 2010. Polarisation measurements have been reported for several gamma-ray bursts. For GRB100826A~\cite{GAP1} the degree of polarisation was found to be (27$\pm$11)\% with the polarisation angle varying during the burst - an observation which, if confirmed, will help to constrain proposed emission mechanisms by probing physical changes in the jet outflow. For GRB110301A and GRB110721A~\cite{GAP2}, a polarisation degree of (70$\pm$22)\% and (84$^{+16}_{-28}$)\%, respectively was reported. For these bursts, no significant change of polarisation angle was seen. Models based on synchrotron emission are consistent with all three GRBs.

This paper describes PoGOLite~\cite{pogolite} - a balloon-borne hard X-ray (25 - $\sim$100 keV) polarimeter which is optimised for the study of compact astrophysical objects. The PoGOLite Collaboration comprises groups from Sweden (KTH Royal Institute of Technology and Stockholm University), Japan (Hiroshima University, Tokyo Institute of Technology, ISAS/JAXA, Nagoya University and Waseda University) and USA (SLAC/KiPAC and University of Hawaii). Affiliated industrial partners include DST Control, Sweden, who have developed the attitude control system and SSC Esrange, Sweden, who have developed the gondola, power and balloon housekeeping systems and are responsible for balloon launches from the Esrange Space Centre.
\section{Polarimeter design}

\begin{figure}[!t]
\centering
\includegraphics[width=9cm]{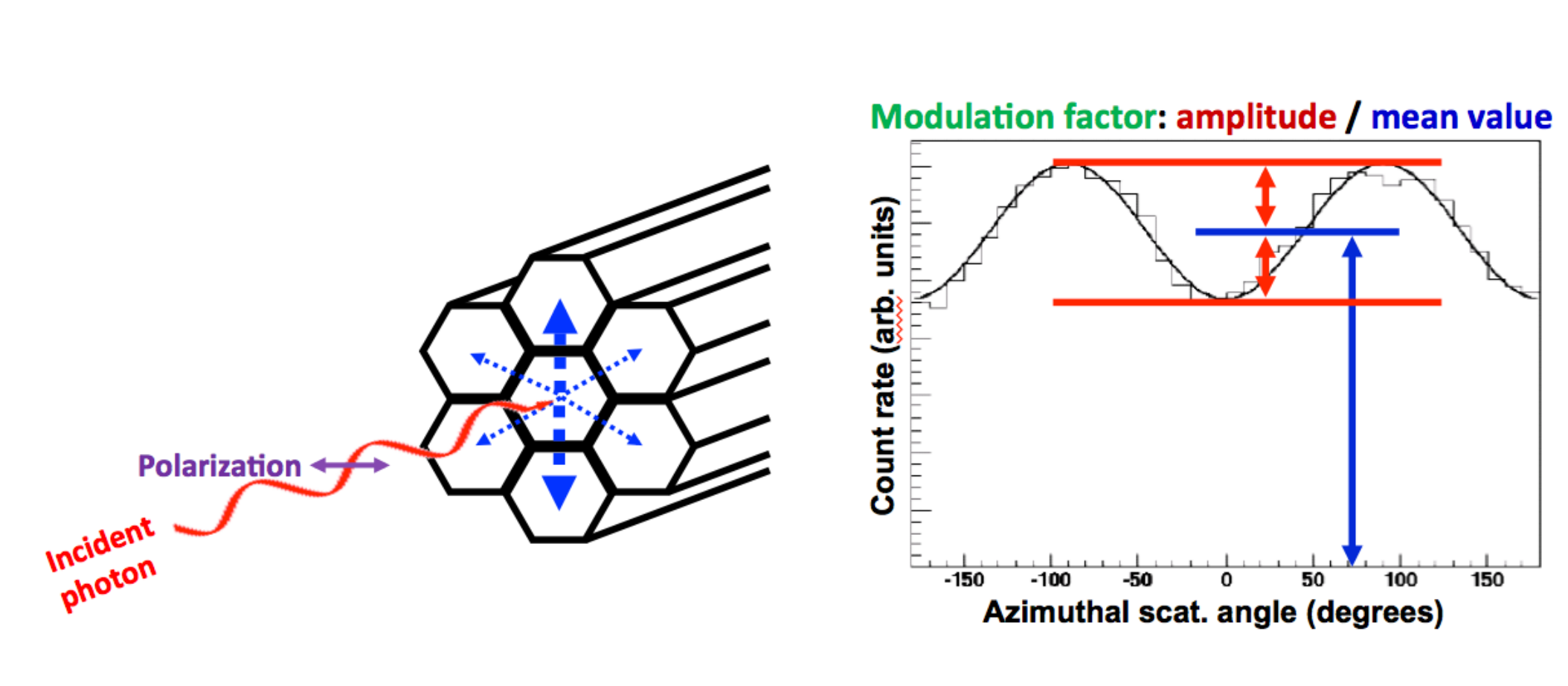}
\caption{The azimuthal scattering angles for a beam of photons will be modulated by the polarisation of the beam. The scattering angle can be reconstructed in a suitably segmented detector (left panel). The resulting modulation curve (right panel) can be used to determine the polarisation of the beam.}
\label{fig:modulation}
\end{figure}

When photons interact in matter, there is a characteristic angle which is modulated by the polarisation of the beam. For photons interacting through the photoelectric effect (typically, $<$10~keV), the azimuthal angle of the emitted photoelectron is modulated. In the case of Compton scattering (typically, 10's to 100's of keV), the azimuthal scattering angle of the photon is instead modulated~\cite{Lei}. For pair production ($>$1~MeV), the modulated quantity is the azimuthal plane containing the electron-positron pair. The degree of modulation is quantified using the modulation factor which is defined in figure~\ref{fig:modulation}. Once the modulation curve is reconstructed, the polarisation degree of the source is defined as the observed modulation factor divided by the modulation factor for a 100\% polarised beam (often called M$_{100}$). This quantity is usually derived from computer simulations validated with polarised photon beam data, e.g. at a synchrotron facility. The phase of the fitted modulation curve yields the polarisation angle.

\begin{figure}[!t]
\centering
\includegraphics[width=9cm]{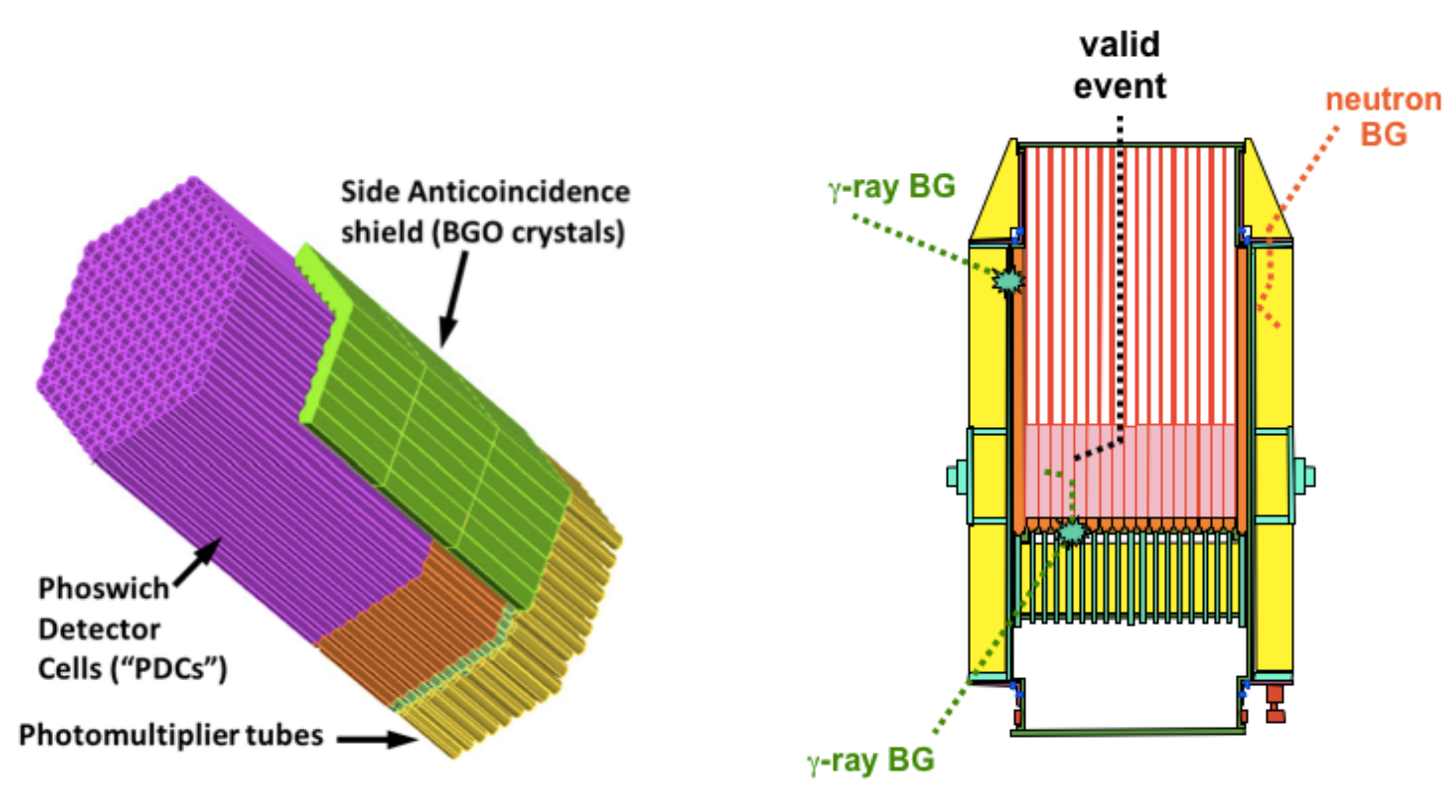}
\caption{Left: an overview of the PoGOLite polarimeter design showing the scintillator components. Right: a schematic cross section of the PoGOLite instrument (not to scale) showing valid and background photon interactions. Possible background atmospheric neutron interactions are also shown. Each PDC is $\sim$1~m long. The side anticoincidence shield (SAS) is segmented to allow background asymmetries to be studied in flight.}
\label{fig:scheme}
\end{figure}

In the case of PoGOLite, polarisation is determined by studying the distribution of azimuthal Compton scattering angles in a segmented detector comprising plastic scintillators which due to their low atomic number provide a favourable cross-section for Compton scattering while also maintaining acceptable sensitivity for photoelectric absorption. Given the constraints of a balloon-borne mission, a plastic-based instrument also allows for a large effective area while maintaining a reasonable M$_{100}$. Polarisation events are characterised by a Compton scatter and a photoelectric absorption in nearby detector elements. The Compton scattering and photoabsorption events are identified in an array of phoswich detector cells (PDC) made of plastic and BGO scintillators, surrounded by a segmented BGO side anticoincidence shield (SAS), as shown in figure~\ref{fig:scheme}. The full-size PoGOLite instrument consists of 217 PDC units. For the maiden 'pathfinder' flight, a reduced volume instrument will be flown consisting of 61 PDC units. 
Each PDC is composed of a thin-walled tube (well) of slow plastic scintillator at the top (fluorescence decay time $\sim$280 ns, length 60~cm), a solid rod of fast plastic scintillator (decay time $\sim$2 ns, length 20~cm), and a BGO crystal at the bottom (decay time $\sim$300 ns, length 4~cm), all viewed by one photomultiplier tube (PMT) - Hamamatsu R7899EGKNP. The wells serve as a charged particle anticoincidence, the fast scintillator rods as photon detectors, and the bottom BGOs act as a lower anticoincidence. Each well is sheathed in thin layers of tin and lead foils to provide passive collimation. 

\begin{figure}[!t]
\centering
\includegraphics[width=9cm]{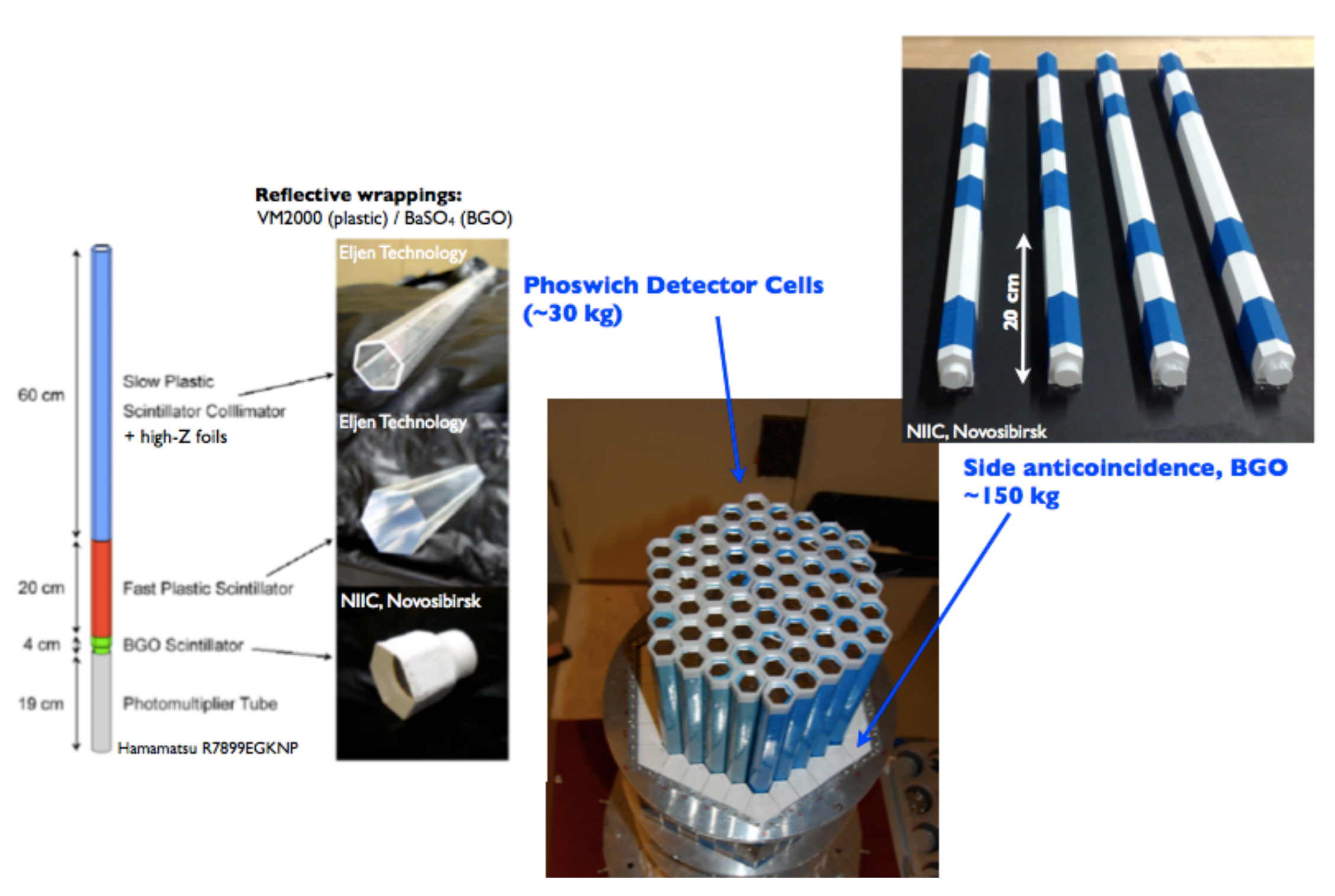}
\caption{The scintillator components of the phoswich detector cells (PDCs) and side anticoincidence system.}
\label{fig:detector}
\end{figure}

Figure~\ref{fig:scheme} shows a simplified cross-section of the polarimeter with possible photon interactions indicated. X-rays entering within the field-of-view of the instrument ($\sim$2$^\circ \times$2$^\circ$ (FWHM), defined by the slow plastic collimators) will hit one of the fast plastic scintillators and may be Compton scattered, with a probability that depends on the photon energy. The scattered photon may escape, be photoabsorbed in another detector, or undergo a second scattering. Electrons resulting from a photoabsorption will deposit their energy in the plastic scintillator and produce a signal at the PMT. A 25~keV Compton scattering event will result in a 1-3~keV energy deposit in the fast plastic scintillator, requiring single photoelectron detection. The PMT is designed to have $\sim$0.05 photoelectron ripple for a gain of 10$^6$. The detection of an energy deposit compatible with photoabsorption initiates high speed waveform sampling of PMT outputs from all PDCs with signals above a threshold. Valid Compton scattering events will be selected from these waveforms after the completion of a flight.  A SpaceWire-based data acquisition system performs the waveform digitisation using 12~bit flash ADCs operating at 37.5~MHz~\cite{daq}. In the absence of veto signals, e.g. from pulse shape discrimination (indicating slow scintillator or anticoincidence activity) or corresponding to the large energy deposits from an interacting charged cosmic-ray, PDC waveforms are saved with 15 pre- and 35 post-trigger samples. The locations of the PDCs in which the Compton scatter and photoabsorption are detected determine the azimuthal Compton scattering angle. The geometry of the PDC arrangement limits the polar scattering angle to approximately (90$\pm$30)$^\circ$, roughly orthogonal to the incident direction. Little of the energy of an incident gamma-ray photon is lost at the Compton scattering site(s). Most of the energy is deposited at the photoabsorption site. This makes it straight-forward to differentiate Compton scattering sites from photoabsorption sites despite the relatively poor energy resolution of plastic scintillator. Each polarisation event can be time-tagged with a precision of $\sim$1$\mu$s using a local precision oscillator synchronised to GPS pulse-per-second information from the attitude control system.

The performance of the polarimeter is discussed in detail elsewhere~\cite{pogolite,performance}. At 50~keV the effective area is 200~cm$^2$ (40~cm$^2$ for the pathfinder) for a M$_{100}$ of $\sim$30\%. The minimum detectable polarisation is expected to be $\sim$10\% for a 200 mCrab (1~Crab for the pathfinder) Crab-like source during a single ballooning campaign. Background contributions to a Crab observation by the PoGOLite pathfinder simulated using Geant4 indicate that the minimum ionising particle signature in scintillators arising from charged cosmic-ray backgrounds is readily reduced using a simple pulse height analysis. The photon background arises from both atmospheric and galactic sources. This background is suppressed due to the narrow field-of-view provided by the PDC design combined with the segmented anticoincidence system. A background due to neutrons (mostly albedo) generated by cosmic-ray interactions with the Earth's atmosphere dominates. Neutrons in the energy range $\sim$0.5 - 50~MeV are the major contributor to the false trigger rate. For this reason a polyethylene shield (with thickness 15~cm around the scattering scintillators) surrounds the polarimeter. This reduces the neutron background by an order of magnitude. The atmospheric neutron energy spectra is poorly measured at stratospheric balloon altitudes. Expectations published in 1973~\cite{Armstrong} for solar minimum conditions at a latitude of 42$^\circ$N provided the original background estimations for PoGOLite with a simple scale factor applied to account for the higher latitude of Kiruna (68$^\circ$N) where PoGOLite balloon flights take place. This work has been revisited within the framework of Planetocosmics~\cite{Planetocosmics, Merlin} - a Geant4-based tool which allows particle interactions with the Earth's atmosphere and magnetosphere to be simulated. An increased neutron background rate was found due to the lower geomagnetic cut-off at Kiruna (although this will be partially offset by present levels of increased solar activity). Data analysis strategies are being studied to mitigate this predicted increase in background. For example, the outer layers of PDCs can be included in the anticoincidence condition. In order to measure the neutron background during flight and directly validate simulation models, a dedicated neutron detector comprising a 5~mm thick LiCaAlF$_6$ crystal with 2\% Eu doping complemented with BGO anticoincidence is mounted in the vicinity of the fast scattering scintillators~\cite{licaf}. The $^{6}$Li in the crystal has a large neutron capture cross-section (940 barn) and is sensitive to neutrons with energies $<$10~MeV. 

The polarimeter is housed inside a pressure vessel system along with the data acquisition electronics. X-rays enter the polarimeter through a thin PEEK window. A fluid-based cooling system is employed to manage the heat load ($\sim$200~W) generated inside the polarimeter. The heat originates from the DC/DC converters inside the tightly-packed array of 92 photomultipliers and from the data acquisition electronics. Fluid (Paratherm LR) is pumped around a cooling plate to which the PMTs are thermally bonded, and around the walls of the housing for the data acquisition electronics. Heat is dissipated through radiators which are directed  towards the cold sky during the flight. An external cooler can be connected during on-ground tests or dry ice placed onto the radiator. It is particularly important that the PMTs are operated close to room temperature in order to assure sensitivity to the small energy deposits expected from photons Compton scattering at the lower end of the PoGOLite energy range. 

The polarimeter pressure vessel is located in a so-called Rotation Frame Assembly (RFA) which allows the polarimeter to rotate around the viewing axis. During a typical observation run of approximately 5~minutes, the polarimeter executes one revolution driven by a compact torque motor mounted on the RFA. Each polarisation event is tagged with the roll angle. Two star trackers  which form part of the attitude control system and an auroral monitor~\cite{auroral} are mounted on the underside of the RFA. Pockets in the RFA house polyethylene blocks which are used to mitigate the neutron background. The RFA also provides a mechanical interface to the elevation torque motors described in the next section. The Payload Control Unit (PCU) is mounted on the upper-side of the RFA, thereby acting as a counter-weight for the star trackers and auroral monitor and ensuring that the polarimeter is well balanced in elevation so that the torque motors can operate efficiently. The PCU houses the power supply for the polarimeter; a redundant PC104-based computer system and a RAID solid-state disk array which controls the polarimeter, collects housekeeping data from (e.g.) temperature and pressure sensors, archives scientific and housekeeping data and commands the attitude control system; Iridium modems for over-the-horizon communications; and interface electronics to the real-time computer system of the ACS. Systems with time-critical functions are typically connected with RS422/485 links, otherwise Ethernet is used. The operating modes and control philosophy for PoGOLite are described elsewhere~\cite{miranda}. The polarimeter and housekeeping data are archived to a robust 'black box' data storage unit which is securely fastened inside the gondola. The black box computer (PC104 and RAID solid-state disk array) can be used to control the entire payload if the primary PCU computer system fails. 

\section{Attitude control system}

The attitude control system (ACS) is designed to keep the viewing axis of the polarimeter aligned to the sidereal motion of observation targets and also compensates for local perturbations such as flight-train torsion and stratospheric winds. Pointing to within $\sim$5\% of the polarimeter field-of-view (corresponding to $\sim$0.1$^\circ$) is required to secure a minimum detectable polarisation of better than 10\% for a 1 Crab source during the pathfinder flight. 

\begin{figure}[!t]
\centering
\includegraphics[width=9cm]{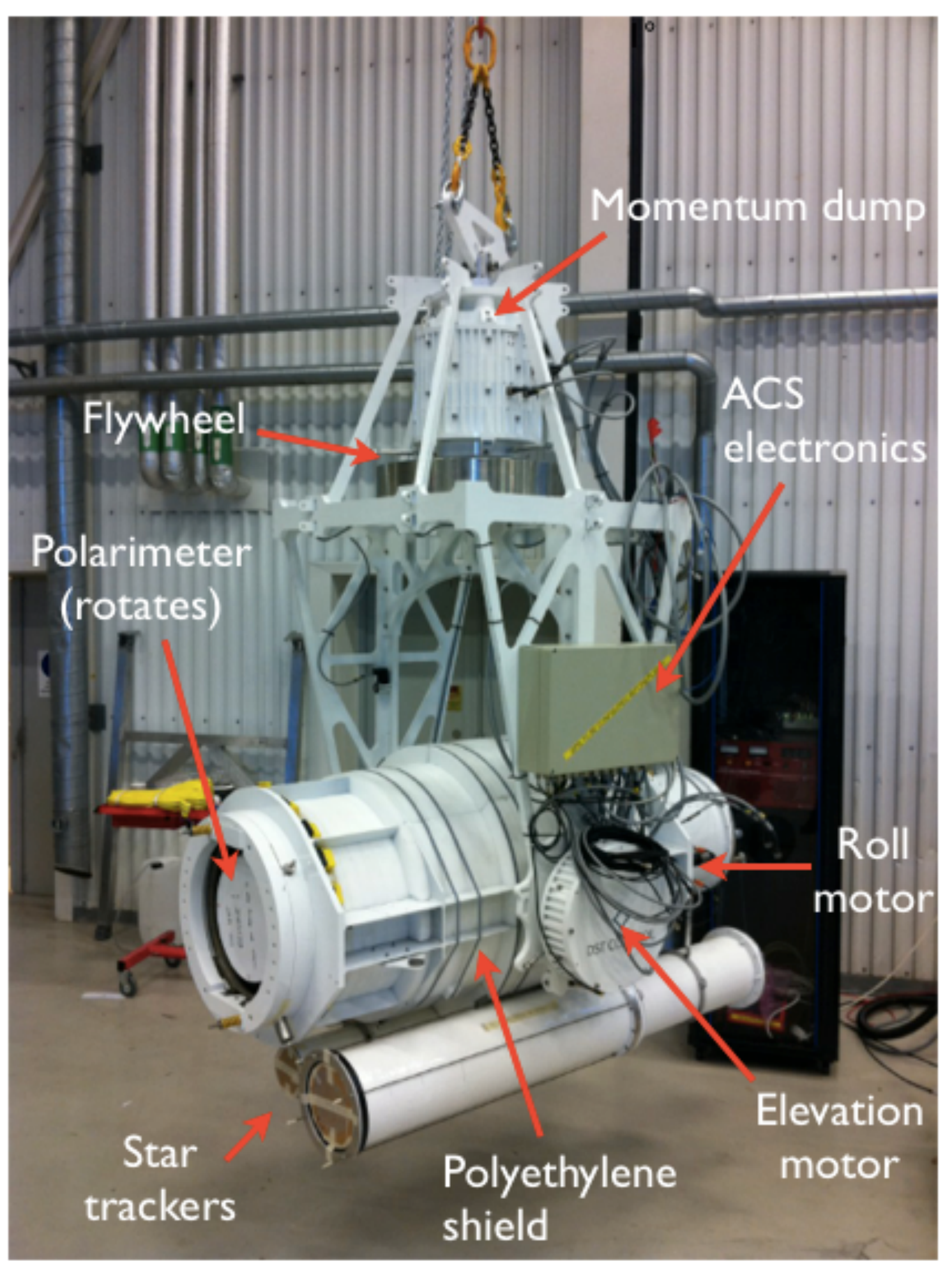}
\caption{The PoGOLite attitude control system. The polarimeter is mounted in the Gimbal Unit Assembly (GUA) where custom direct drive torque motors act on the polarimeter elevation axle of the Rotation Frame Assembly (RFA). The polarimeter can roll around the viewing axis within the RFA. The balloon flight train connects to the gimbal assembly through a combined momentum dump/flywheel system which is used for azimuthal positioning. The Payload Control Unit (PCU) and auroral monitor are not shown.}
\label{fig:acs}
\end{figure}

\begin{figure}[!t]
\centering
\includegraphics[width=9cm]{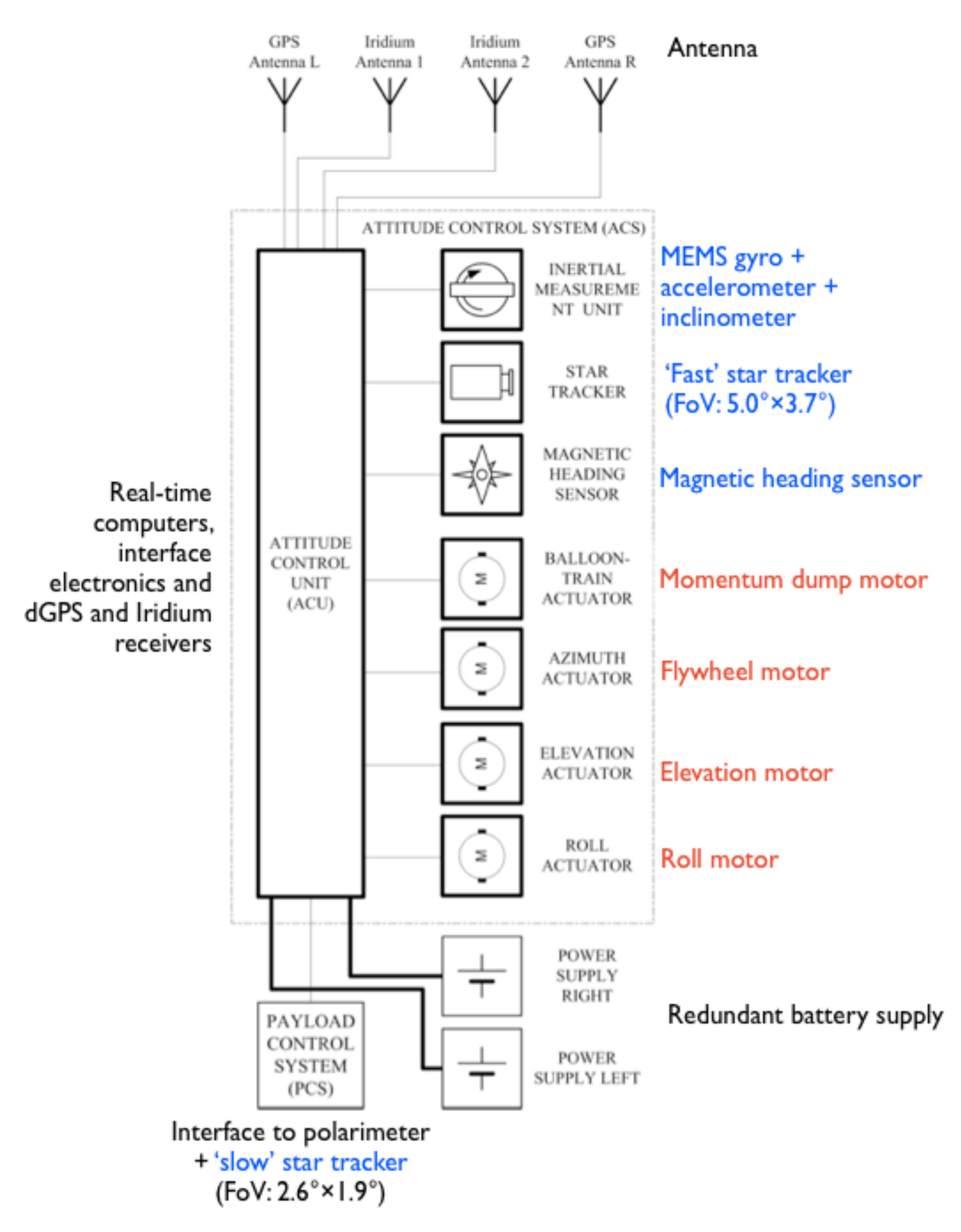}
\caption{A simplified block diagram of the PoGOLite attitude control system.}
\label{fig:acs_schem}
\end{figure}

The attitude control system has been developed by DST Control in Link\"{o}ping, Sweden~\cite{DST}. The components used are free from ITAR restrictions. The polarimeter and RFA are mounted in the ACS Gimbal Unit Assembly (GUA), as shown in figure~\ref{fig:acs}. Custom torque motors~\cite{jes} act directly on the polarimeter elevation axis. Azimuthal positioning is achieved with a flywheel assembly which connects to the flight train through a momentum dump motor, allowing angular momentum stored in the flywheel to be reset upon saturation and to null out flight train bearing friction. Control signals to the motor systems are generated by a real-time computer system which monitors the attitude sensors, comprising a differential GPS system, a 3-axis micromechanical accelerometer/gyroscope package, angular encoders (elevation, flywheel, momentum dump and instrument roll), an inclinometer and a 3-axis magnetometer. A schematic overview is presented in figure~\ref{fig:acs_schem}. These attitude sensors are augmented by two star trackers. One star tracker is developed from a design successfully used on previous missions~\cite{stm} and has a field-of-view of 2.6$^\circ \times $1.9$^\circ$. Ground-based tests~\cite{cmb} have shown that stars down to 10$^{\mathrm{th}}$ magnitude can be resolved assuming the background light conditions at 40~km~\cite{cmb91} (for an August turn-around flight). The second star tracker is a more compact design with in-house designed optics providing a field-of-view of 5.0$^\circ \times $3.7$^\circ$. Both trackers require a stabilized gondola to operate (provided by the differential GPS system and gyroscopes) and can either operate in a star pattern matching mode (providing relatively slow, but absolute position fixes) or can be commanded to lock onto a bright star providing ACS updates at a rate of 10-100~Hz. The latter case is foreseen to be the primary operating mode during flight. 

\section{Communications}

When in line-of-sight (LOS) from Esrange, communications with the gondola use the E-Link system developed by SSC~\cite{Elink}. This is an implementation of 10/100 Base-T Ethernet over a S-band radio system which supports duplex traffic at up to 2 Mb/s. The maximum LOS range is approximately 500~km for a balloon at float altitude. Atmospheric conditions may have a significant impact on the link range. An updated version of E-Link with a data capacity exceeding 20 Mb/s (with lower mass and power consumption) is under development and will be flown on future PoGOLite flights. Once the gondola drops below the horizon, the Iridium satellite communications system is used. Two (for redundancy) dial-up data links are used to monitor and potentially control the ACS and two RUDICS (Router-based Unrestricted Digital Internetworking Connectivity Solutions) links are used to monitor and command the entire payload. The dial-up links operate in a point-to-point mode, with Iridium modems and antenna on the ground used to dial-up to modems on-balloon (via the Iridium satellite constellation). For the RUDICS links, modems on-balloon communicate directly with Iridium's commercial ground-stations (again, via the Iridium satellite constellation) where the call is terminated at a user specified IP address.  For PoGOLite, this corresponds to a redundant server system located in Stockholm. This system has the advantage that TCP/IP packets can be sent directly over the link. On-balloon, the dial-up and RUDICS links are cross-connected to provide an additional layer of redundancy in case of link failure. The data rate is far inferior to E-Link at approximately 1~kb/s per link. Reliable on-board data storage and space efficient housekeeping data is therefore vital. During the Iridium phase of the flight, the gondola is foreseen to operate autonomously through a predefined observation schedule. There are additional Iridium links (dial-up and SBD - 'short burst data') for Esrange housekeeping and command functions.

\section{Gondola and ancillary systems}

\begin{figure}[!t]
\centering
\includegraphics[width=9cm]{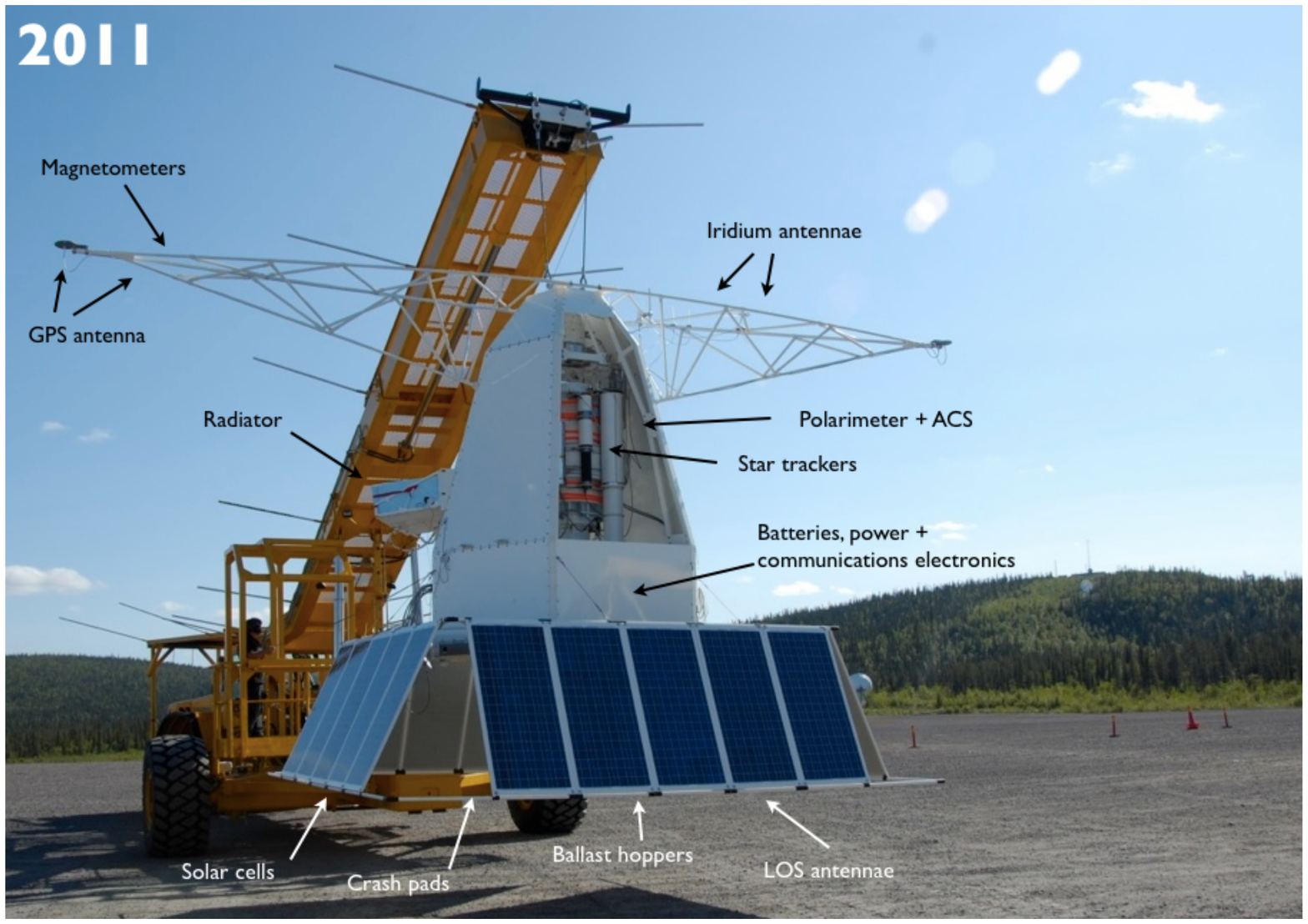}
\caption{The complete gondola assembly ready for flight - 2011 configuration.}
\label{fig:gondola2011}
\end{figure}

As shown in figure~\ref{fig:gondola2011}, the polarimeter and ACS are installed in a gondola assembly (designed by SSC Esrange). The upper part of the gondola is connected to the Gimbal Unit Assembly and also provides a mounting point for the cooling system radiator and pump. The lower section houses batteries, power control electronics and communications equipment. The gondola frame is covered in lightweight composite honeycomb panels which enhance the structural rigidity, help protect the polarimeter from damage during landing and provide thermal shielding. In the 2011 configuration, two glass-fibre booms were attached to the top of the gondola. The booms span $\sim$10~m and have a GPS antenna mounted at each end. The antenna separation provides the required baseline for the differential GPS system alone to meet the $\sim$0.1$^\circ$ pointing accuracy requirement. Other GPS antennas for Esrange flight systems, Iridium antenna and magnetometers for the ACS and auroral monitor are also mounted on the booms. Beneath the lower gondola, a four-sided 'skirt' of solar panels (each side comprising 5 panels, with overall dimensions (3.5 $\times$ 1.5)~m was mounted along with landing crash pads, ballast hoppers and E-Link communication antennas. The gondola stands approximately 5~m tall.

\section{Balloon campaigns and outlook}

The PoGOLite balloon campaigns are conducted at the Esrange Space Centre, outside of Kiruna, in the North of Sweden (68$^\circ$N, 21$^\circ$E)~\cite{stig}. PoGOLite is carried into the stratosphere by a helium-filled polyethylene balloon. The balloon has a volume of 1.1$\times$10$^6$~m$^3$ and an average thickness of $\sim$20~$\mu$m. Launches typically occur early in the morning when surface wind velocities are low. The ascent to float altitude takes approximately 2.5~hours. Once at float, the balloon will remain over the launch site during so-called turn-around periods (early May and latter half of August) or is carried in a Westerly direction by the stratospheric winds during mid-May to end-July. Launches during the Winter period generally give short flight times owing to strong stratospheric winds and recovery restrictions. The launch window for PoGOLite opens on July 1st and closes approximately 1~month later. This period is chosen in order to satisfy three conditions:
\begin{enumerate}
\item A Westerly trajectory allows a long duration flight with termination on Victoria Island, Canada, after $\sim$5~days or after a circumpolar traverse of the North Pole, after $\sim$15~days. 
\item After 1st July the angle between the Crab and the Sun exceeds 15$^\circ$ which is a requirement from the star tracking baffle design\footnote{The PoGOLite ACS was originally designed for a turn-around flight. Future PoGOLite flights could benefit from the inclusion of a Sun tracker.}.
\item Around the end of July, the Westerly stratospheric winds begin to decrease heralding the onset of the August turn-around period.
\end{enumerate}

The total gondola mass is $\sim$1750~kg. Several hundred kilograms of ballast are also flown in order to give the balloon pilot control during the ascent and termination phases of the flight. In comparison, the balloon material weighs $\sim$2000~kg, the parachute $\sim$270~kg, and the rigging $\sim$120~kg. An initial float altitude exceeding 38.5~km is foreseen, corresponding to an overburden of less than $\sim$4~g/cm$^2$. The float altitude follows an diurnal pattern with the altitude decreasing when the Sun is low (or under the horizon) leading to a decrease in the helium gas temperature. Conversely, the effect of solar heating causes the float altitude to increase. The combination of a valve at the top of the balloon and ballast drops can be used to stabilise the balloon altitude. The primary observation goals of the pathfinder flight are the Crab and Cygnus X-1. During times when neither target is visible, background studies will be performed.

\begin{figure}[!t]
\centering
\includegraphics[width=9cm]{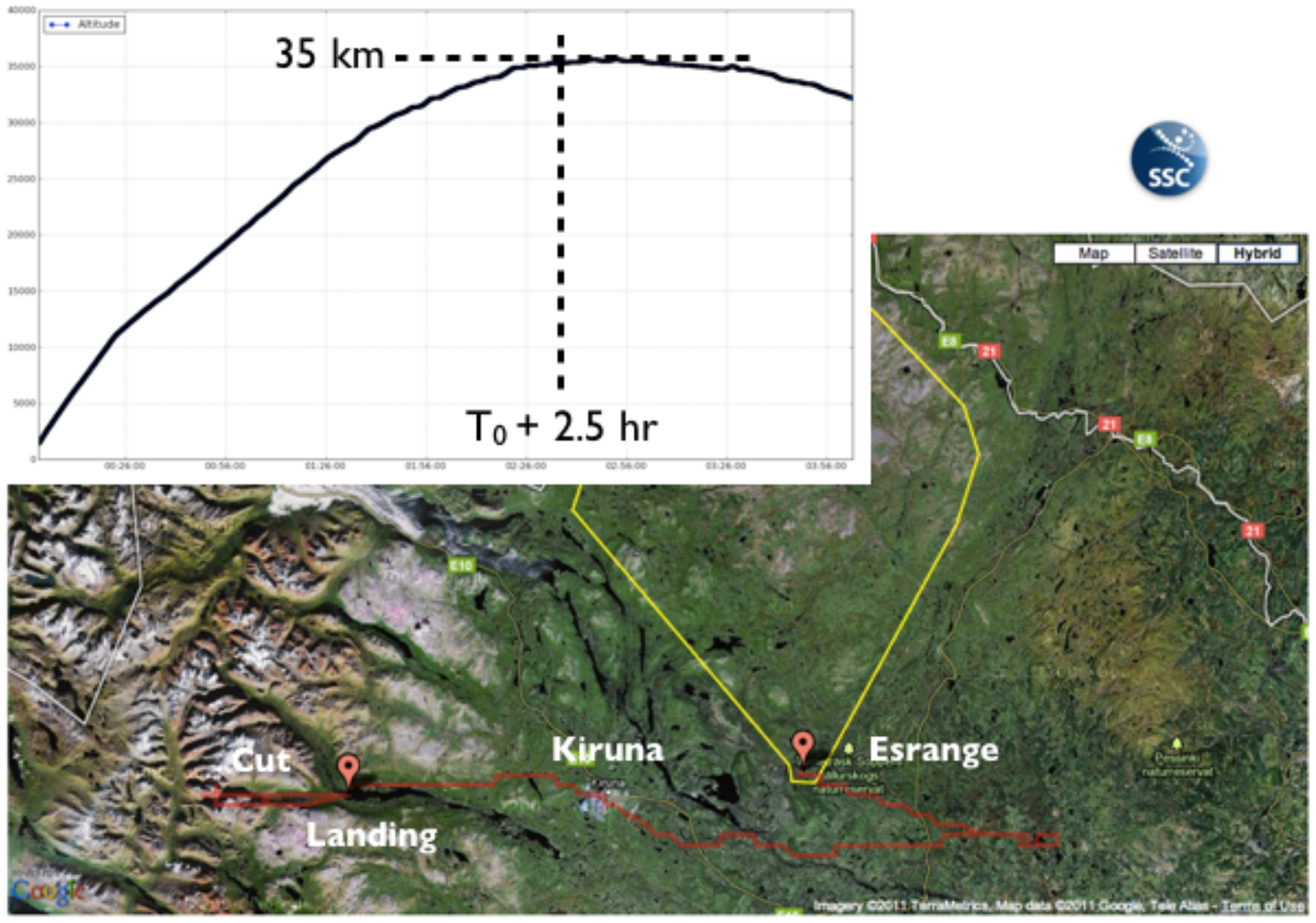}
\caption{The ascent profile of the 2011 PoGOLite flight.}
\label{fig:ascent}
\end{figure}

The PoGOLite pathfinder mission was launched from the Esrange Space Centre at $\sim$02:00 local time on July 7th 2011. The ascent profile is shown in figure~\ref{fig:ascent}. At $\sim$04:30 the altitude levelled out at 35 km and at $\sim$05:20 the altitude started to decrease due to a suspected balloon leak. The decision to terminate the flight was taken soon after. At $\sim$06:00 PoGOLite systems were turned off and at $\sim$07:20 the balloon was cut from the gondola close to the highest mountain in Sweden, Kebnekaise. The gondola touched down by parachute at $\sim$08:01 and was recovered close to the village of Nikkaluokta and returned to Esrange. A post-flight investigation indicated that the balloon failure was due to a sudden change in the surface wind direction and increase in wind speeds at the time of launch. Relatively minor damage to the gondola was found. However, on-board accelerometers indicated significantly higher shocks ($\sim$25 g peak) upon landing than had been anticipated and mitigated through design choices (10 g peak). The emergency termination of the flight before encountering mountainous terrain on the Norwegian border resulted in a landing among piles of rocks close to Laukkuj\"{a}rvi lake. The landing took place in a valley, resulting in relatively large transverse wind velocities and a resulting relatively high impact velocity. As a precaution, the polarimeter was completely disassembled. A small number of broken photomultiplier tube windows and plastic scintillator glue joints were identified and replaced using available spares. No significant structural damage was found and all electronics systems worked as expected. In parallel with this work, the attitude control system was disassembled at DST Control. No major damage was found but all bearing assemblies were exchanged as a precaution. Although somewhat unwelcome, the premature flight termination provided an opportunity to validate the mechanical design of PoGOLite and confirm that the payload design would allow multiple flights with only minor repairs required in between.

\begin{figure}[!t]
\centering
\includegraphics[width=9cm]{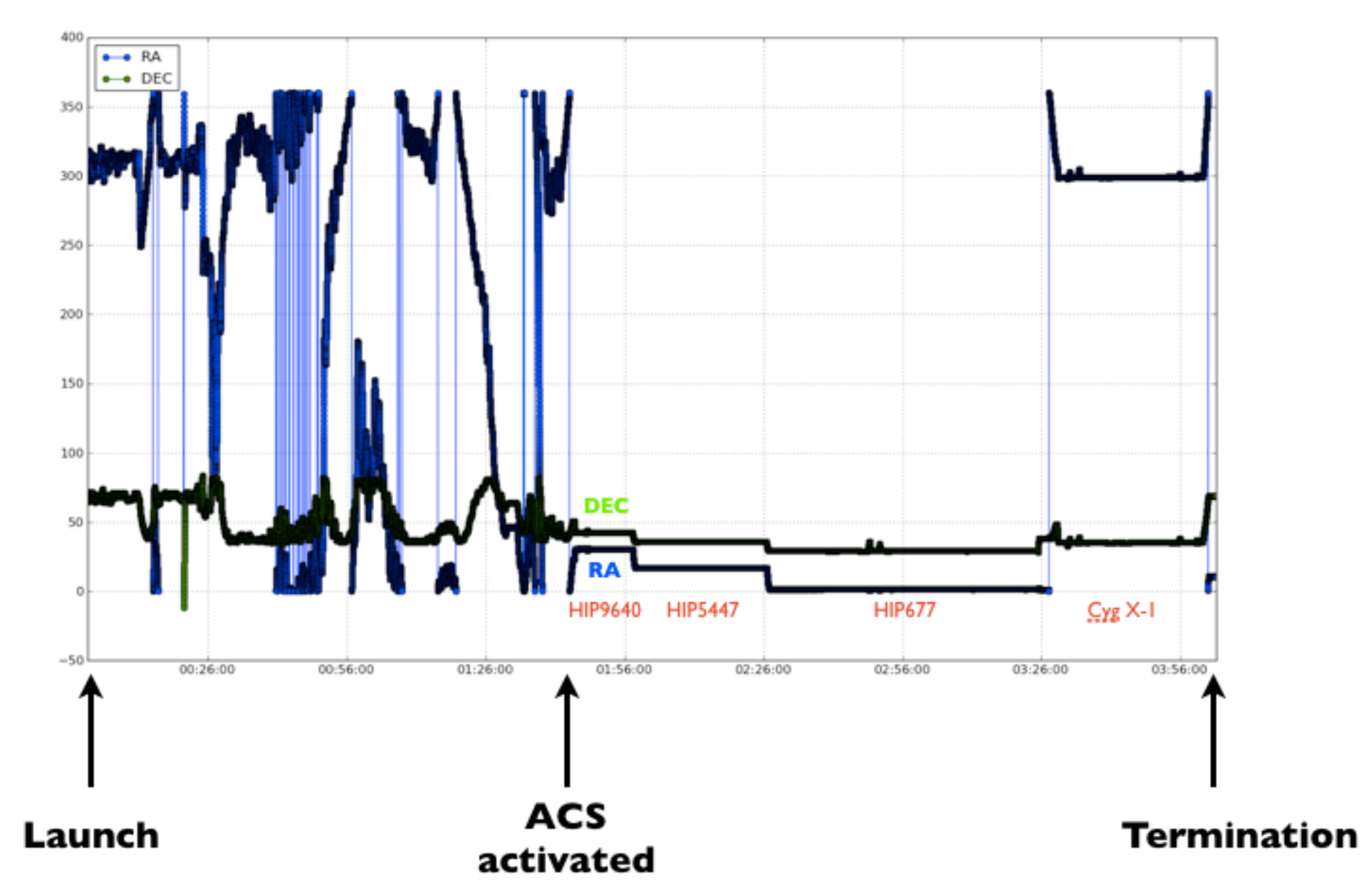}
\caption{The timeline of the maiden flight of the PoGOLite pathfinder in 2011. The reconstructed RA and Dec pointing directions are displayed.}
\label{fig:acs_perf1}
\end{figure}

\begin{figure}[!t]
\centering
\includegraphics[width=9cm]{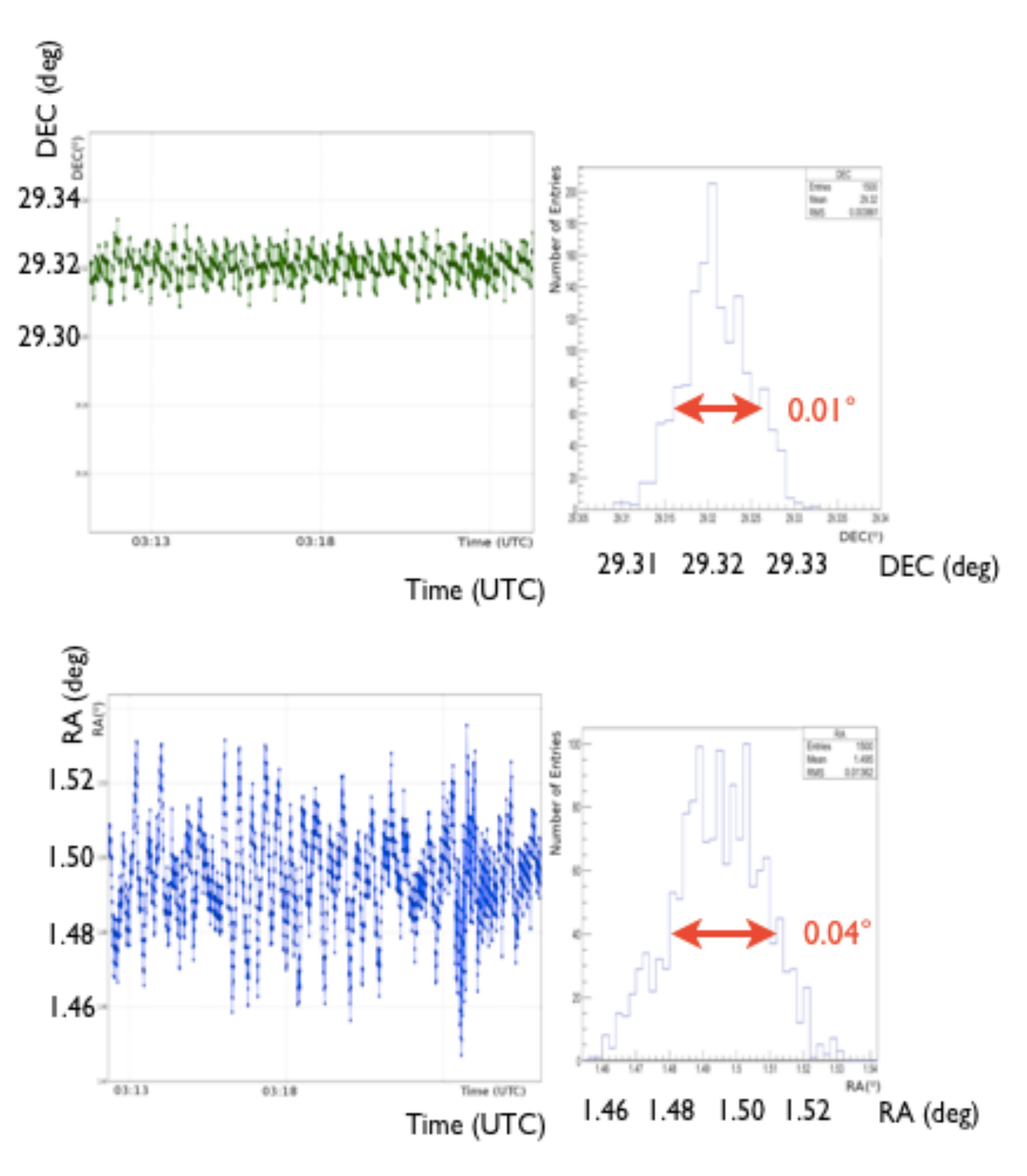}
\caption{The distribution of reconstructed RA and Dec values during part of a Cygnus X-1 observation.}
\label{fig:acs_perf2}
\end{figure}

The relatively low float altitude attained prior to termination, coupled with the short time at float precluded any scientific measurements. A number of basic performance tests of the polarimeter and attitude control system were however possible~\cite{Merlin}. As an example, performance studies of the ACS are outlined here. The timeline for ACS commissioning is shown in figure~\ref{fig:acs_perf1}. During approximately the first 1.5~hours of ascent (up to an altitude of $\sim$25~km), the ACS was not activated and the gondola was left to rotate freely under the balloon. This motion provided uniform solar illumination of the gondola and helped to prevent possible freezing of the momentum dump bearings during passage through the tropopause where temperatures were observed to fall below -45$^\circ$C (by comparison, the temperature at ceiling, $\sim$35~km, was  -10$^\circ$C). For this reason, the flywheel, elevation and roll motors were put into 'exercise' mode so that the there was a continuous small oscillatory movement on the motor axes. Once activated, commissioning of the ACS proceeded initially by tuning ACS control loop parameters to account for the long flight train rigging which was impractical to use during ground-based tests. After a first tuning pass, the ACS was commanded to point at a series of guide stars and offsets in the star tracking systems were studied. During this procedure, notification of flight termination arrived. None-the-less, the performance of the ACS after a preliminary tuning was found to be very promising, as shown in figure~\ref{fig:acs_perf2}. This data shows a representative 15~minute long study of the stability of the reconstructed Right Ascension (RA) and Declination (Dec) while tracking on the guide star foreseen for observations of Cygnus X-1. The FWHM of the RA (Dec) distributions was found to be 0.04$^\circ$ (0.01$^\circ$) which is well within specification. Improved performance is expected once the ACS is fully commissioned in flight.

\begin{figure}[!t]
\centering
\includegraphics[width=9cm]{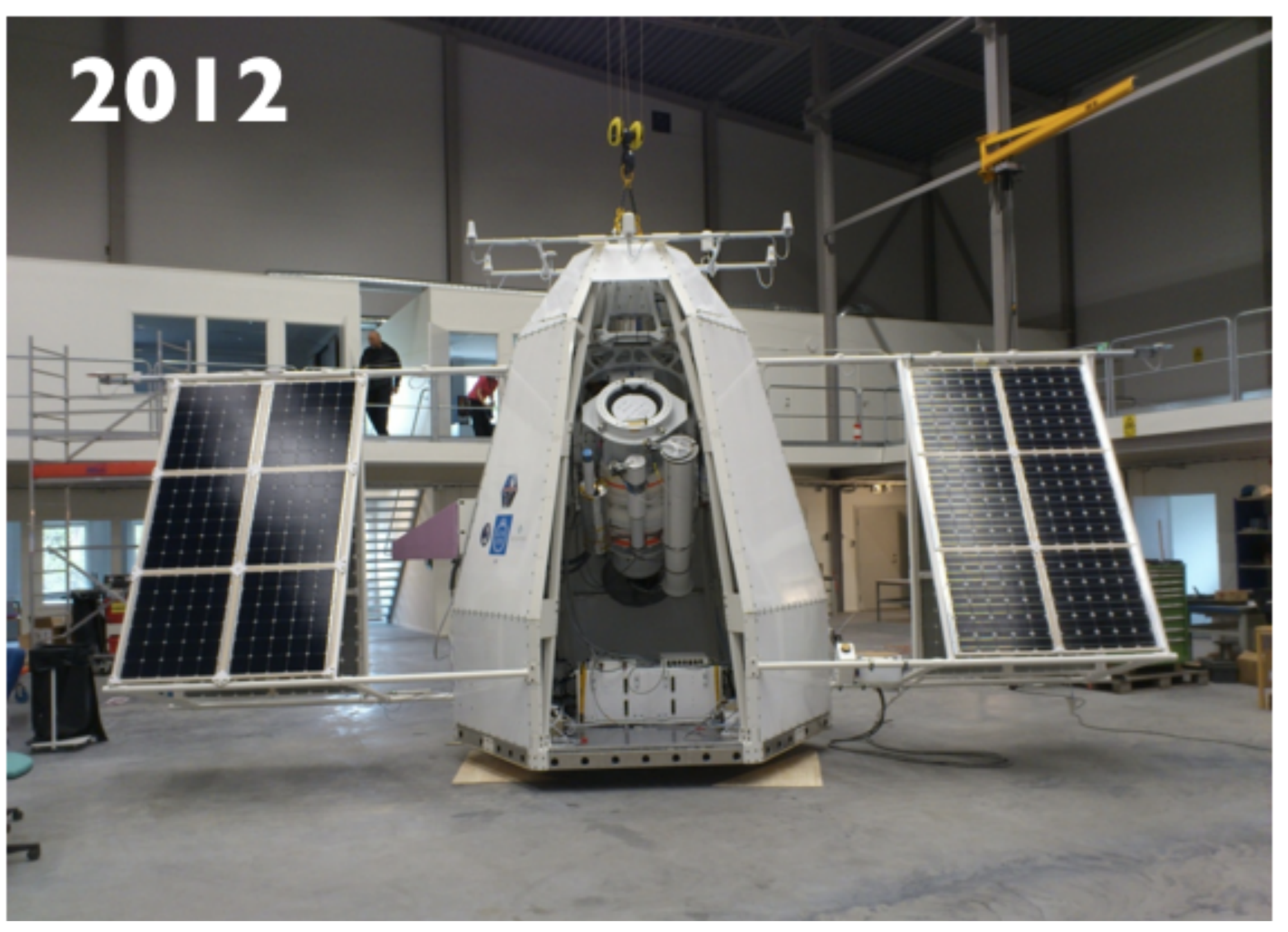}
\caption{The PoGOLite gondola - 2012 configuration.}
\label{fig:newgondola}
\end{figure}

During preparations for the 2012 reflight, the gondola configuration was changed in order to simplify on-ground handling and to reduce the gondola mass. In particular, the solar cell configuration was changed and the power control system optimised, allowing for a reduction in the number of batteries. The new gondola configuration is shown in figure~\ref{fig:newgondola}.

The 2012 launch window opened as planned on July 1st. Frustratingly, weather conditions were never good enough to allow a launch attempt. A low pressure region oscillated back-and-forth over the Esrange region during most of July yielding high surface wind speeds which were incompatible with balloon launch requirements. Based on information regarding the projected evolution of stratospheric wind conditions and near-term local surface weather forecasts, the PoGOLite launch campaign was terminated on July 30th 2012.

PoGOLite is approved for another launch campaign at Esrange in Summer 2013. 

\section*{Acknowledgment}
We acknowledge valuable contributions from undergraduate students Takafumi Kawano, Anders Jonsson, Victor Mikhalev, Maria Mu\~{n}oz-Salinas and Hadrien Verbois; and from our colleagues at DST Control and SSC, Esrange Space Centre. 


\end{document}